\documentclass[%
 reprint,
superscriptaddress,
footinbib,
 amsmath,amssymb,
 aps,
 prx,
 showkeys,
]{revtex4-1}

\usepackage{graphicx}% Include figure files
\usepackage{dcolumn}% Align table columns on decimal point
\usepackage{bm}% bold math
\newcommand{\be}{\begin{equation}}
\newcommand{\ee}{\end{equation}}
\newcommand{\bea}{\begin{eqnarray}}
\newcommand{\eea}{\end{eqnarray}}

\begin{document}

\preprint{APS/123-QED}

\title{The EPR paradox and quantum entanglement at sub-nucleonic scales}% Force line breaks with \\
% \thanks{zhoudunming@bnl.gov}%

\author{Zhoudunming Tu}
 \email{zhoudunming@bnl.gov}
\affiliation{%
 Department of Physics, Brookhaven National Laboratory, Upton, NY 11973, USA
}%

\author{Dmitri E. Kharzeev}
%  \homepage{http://www.Second.institution.edu/~Charlie.Author}
\affiliation{
 RIKEN-BNL Research Center, Brookhaven National Laboratory, Upton, NY 11973, USA
}%
\affiliation{
 Department of Physics and Astronomy, Stony Brook University, New York, 11794, USA
}%
\author{Thomas Ullrich}
\affiliation{%
Department of Physics, Brookhaven National Laboratory, Upton, NY 11973, USA
}%
\affiliation{
 Department of Physics, Yale University, New Haven, CT 06511, USA
}%

% \collaboration{CLEO Collaboration}%\noaffiliation

\date{\today}% It is always \today, today,

\begin{abstract}
In 1935, in a paper~\cite{Einstein:1935rr} entitled ``Can quantum-mechanical description of reality be considered complete?", Einstein, Podolsky, and Rosen (EPR) formulated an apparent paradox of quantum theory. They considered two quantum systems that were initially allowed to interact, and were then later separated. A measurement of a physical observable performed on one system then had to have an immediate effect on the conjugate observable in the other system -- even if the systems were causally disconnected! The authors viewed this as a clear indication of the inconsistency of quantum mechanics. In the parton model~\cite{Bjorken:1969ja,Feynman:1969wa,Gribov:1968fc} of the nucleon formulated by Bjorken, Feynman, and Gribov, the partons (quarks and gluons) are viewed by an external hard probe as independent. The standard argument is that, inside the nucleon boosted to an infinite-momentum frame, 
the parton probed by a virtual photon with virtuality $Q$ is causally disconnected from the rest of the nucleon during the hard interaction. Yet, the parton and the rest of the nucleon have to form a colour-singlet state due to colour confinement, and so have to be in strongly correlated quantum states -- we thus encounter the EPR paradox at the sub-nucleonic scale. In this paper, we propose a resolution of this paradox based on the quantum entanglement of partons. We devise an experimental test of entanglement, and carry it out using data on proton-proton collisions from the Large Hadron Collider (LHC). Our results provide a strong direct indication of quantum entanglement at sub-nucleonic scales. 
% \begin{description}
% Secondary publications and information retrieval purposes.
% You may use the \texttt{description} environment to structure your abstract;
% use the optional argument of the \verb+\item+ command to give the category of each item. 
% \end{description}
\end{abstract}

%Use showkeys class option if keyword
\maketitle

%\tableofcontents

\section{Introduction}

In quantum mechanics, the entanglement of the quantum states of particles implies that a measurement performed on one of the particles affects the state of the entangled particle, even when they are located at large distances. At first glance, this implies that 
information would have to travel faster than the speed-of-light, which is forbidden by special relativity; this was referred by Einstein as ``spooky action at a distance". In recent years, quantum entanglement has become the base of new technology, including quantum computers~\cite{Steane:1997kb} and quantum cryptography~\cite{Hughes:1995us}. At the same time, studies of entanglement in hadron physics are just beginning. 

%%%%%%%%%%%%%%%%%%%%%%%%
The confinement of coloured quarks inside hadrons provides perhaps the most dramatic example of quantum entanglement that exists in nature. Indeed, the quarks within the hadrons are not just correlated, they simply do not exist in the physical spectrum as isolated objects.
The mechanism of colour confinement is one of the most challenging problems in modern physics. It is universally believed that Quantum Chromodynamics (QCD) should correctly describe this phenomenon, but the underlying dynamics is still mysterious. We hope that recasting this problem in the language of quantum information can shed new light on it, and open new venues for experimental investigations, including the one that we describe below.

Information about short-distance QCD and the parton structure is provided by hard processes, such as Deep Inelastic Scattering (DIS) of leptons off nucleons. Hard processes are characterized by a large momentum transfer, $Q$, and probe short transverse distances $\sim 1/Q$ inside the nucleons and nuclei. About fifty years ago, Bjorken, Feynman, and Gribov formulated the ``parton model"~\cite{Bjorken:1969ja,Feynman:1969wa,Gribov:1968fc} of the nucleon that became the base for theoretical description of hard processes. The parton model is usually formulated in a frame in which the nucleon possesses a large momentum, and the partons are viewed by an external hard probe as {\it independent} constituents that carry different fractions $x$ of the nucleon's momentum. The assumption that the quarks and gluons confined inside the nucleon may be viewed as independent is striking. The reasoning that is commonly invoked to justify this assumption is that
the parton probed at a short distance $\sim 1/Q$ is causally disconnected from the rest of the nucleon during the hard interaction.

Nevertheless, in spite of the spectacular success of the parton model, its assumptions raise a number of conceptual questions~\cite{Kharzeev:2017qzs}. Indeed, the hadron in its rest frame is described by a pure quantum state $|\psi\rangle$ with density matrix $\hat{\rho} = |\psi\rangle \langle \psi |$ and zero von Neumann entropy $S = - \rm{tr} \left[ \hat{\rho} \ln \hat{\rho} \right] = 0$. How does this pure state evolve to a set of ``quasi-free" partons in the infinite-momentum frame? If the partons were truly free and thus incoherent, they should be characterized by a non-zero entropy resulting from different positioning of partons in the configuration space. Since a Lorentz boost cannot transform a pure state into a mixed one, what is the precise meaning of ``quasi-free''? What is the rigorous definition of a parton distribution when applied to a pure quantum state?

\section{\label{sec:theory}Entanglement entropy and parton distributions}

\begin{figure}[thb]
\includegraphics[width=3.5in]{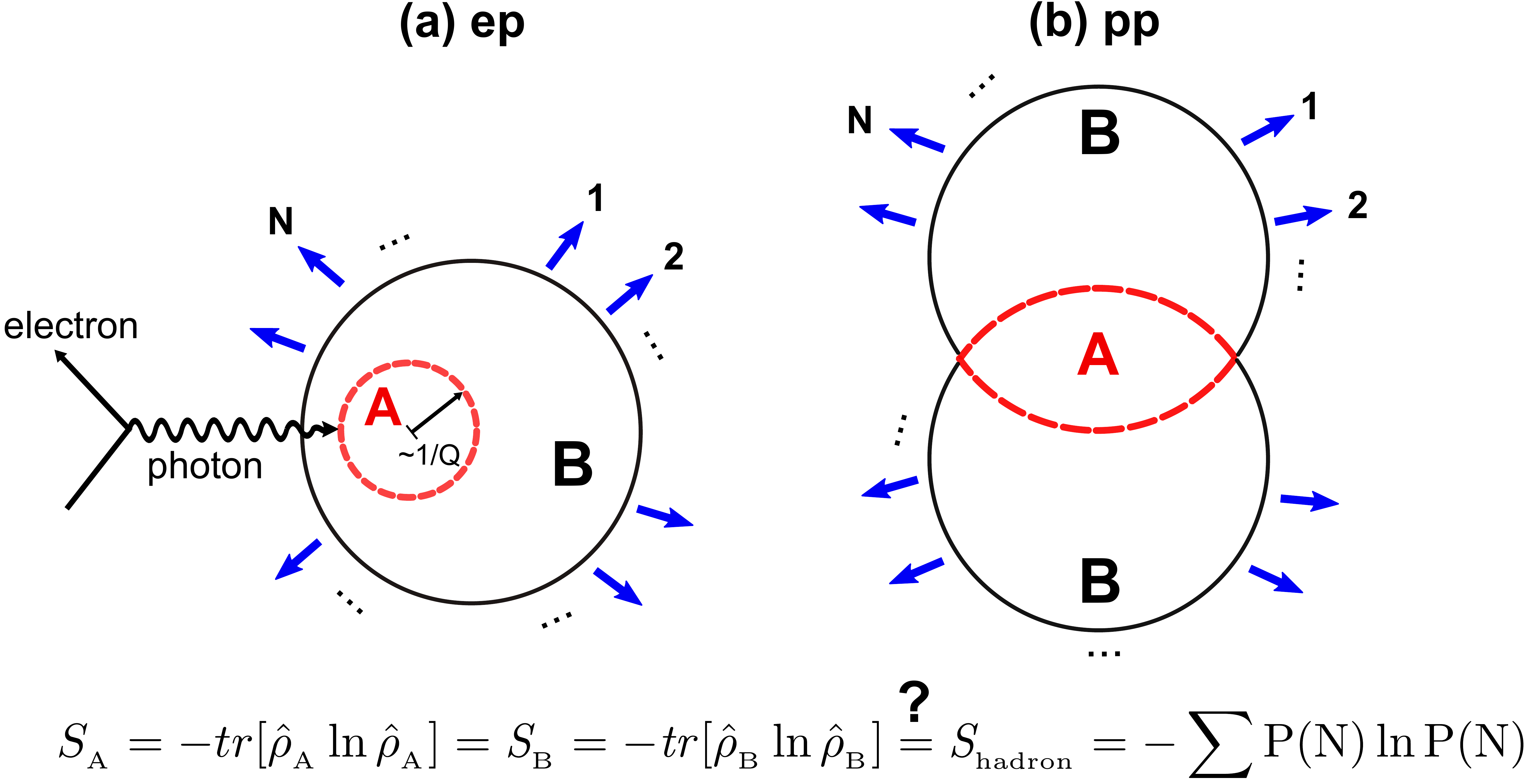}
%   \captionsetup{format=plain,justification=justified, singlelinecheck=false}
  \caption{ \label{fig:figure_1} Illustrations of quantum entanglement in high energy collisions. (a) electron-proton (ep) deeply inelastic scattering, where the virtual photon emitted by the electron probes part of the proton, denoted as region A, while the unobserved part of the proton is represented by region B. (b) proton-proton inelastic collision, where the interaction region is A and the remainder of the system is B. The initial von Neumann entropy from region A and B are denoted as $S_{\mathrm{A}}$ and $S_{\mathrm{B}}$ respectively. The final-state hadron entropy, $S_{\rm hadron}$, is given by the Boltzmann entropy based on the hadron multiplicity distribution $\rm P(N)$. }
\end{figure}

Recently, it has been suggested that this apparent paradox can be resolved by the quantum entanglement of partons~\cite{Kharzeev:2017qzs}. Indeed, consider an electron-proton scattering with momentum transfer $Q$ depicted in Fig.~\ref{fig:figure_1} (a). It is clear that since the transverse distance involved in this process $\sim 1/Q$ is much smaller than the size of the proton, DIS probes only a part of the proton's wave function; let us denote it ${\rm A}$. In the proton's rest frame, where it is definitely described by a pure quantum mechanical state, DIS probes the spatial region ${\rm A}$ localized within a tube of radius $\sim 1/Q$ and length $\sim 1/(mx)$, where $m$ is the proton's mass and $x$ the momentum fraction of the struck quark. Inclusive DIS measurements sum over the unobserved part of the wave function localized in region ${\rm B}$ complementary to ${\rm A}$. Hence we have access only to the reduced density matrix $\hat{\rho}_{\rm A} = \rm{tr}_{\rm B} \hat{\rho}$, but not the entire density matrix $\hat{\rho} = |\psi\rangle \langle \psi |$. The von Neumann entropy, arising from the quantum entanglement between region ${\rm A}$ and ${\rm B}$, namely $S_{\rm A} = - \rm{tr} \left[ \hat{\rho}_{\rm A} \ln \hat{\rho}_{\rm A} \right]$ associated with the DIS measurement, is found~\cite{Kharzeev:2017qzs} to correspond to the entropy of independent partons, and thus to the parton distribution. 
Because the region B is complementary to region A, the entanglement entropy $S_\mathrm{B}$ associated with it
has to be equal to $S_\mathrm{A}$.

At small $x$, where gluons dominate, the relation between the entanglement entropy $S_{\rm A}$ and the gluon distribution\footnote{Probability density for finding a gluon with a certain longitudinal momentum fraction $x$ at resolution scale $Q^2$. } 
$xG(x)$
becomes simply:
\be\label{ent_part}
S_{\rm A} = \ln[ xG(x) ] = S_{\rm B}.
\ee
Here we do not explicitly indicate the dependence of $G$ on the momentum transfer, $Q^2$. Equation (\ref{ent_part}) implies that at small $x$ all microstates of the system are equally probable and the von Neumann entropy is maximal.

In the present paper we devise an independent experimental test of measuring the entanglement entropy of partons within the nucleon using the final-state hadron multiplicity distribution $\rm P(N)$, where $\rm P(N)$ is the probability of producing $\rm N$ particles in the system per event. 
This will allow us to test the proposed relation between the entanglement entropy and the parton distribution given by (\ref{ent_part}).

In DIS experiments, the value of the entropy arising from entanglement depends on the photon probe in terms of $x$ and $Q^{2}$. 
However, the entropy $S_\mathrm{A}$ resulting from the entanglement of region ${\rm A}$ with ${\rm B}$ and giving rise to the parton distribution, should always be equal to the entropy $S_\mathrm{B}$ resulting from the entanglement of region ${\rm B}$ with ${\rm A}$ and giving rise to the final-state entropy of the fragmenting nucleon.
The latter quantity can be reconstructed from the multiplicity distribution of the produced hadrons, which we will denote as 
$S_\mathrm{hadron}$. Based on this argument and the relation (\ref{ent_part}), we thus expect the following relation that can be directly tested in experiments:
\be\label{test}
\ln[ xG(x) ] = S_\mathrm{hadron}.
\ee
This relationship can also be explored in proton-proton ($pp$) collisions, as illustrated in Fig.~\ref{fig:figure_1} (b). In this case, the interaction region of the two protons is the region A, whereas the remaining system is region B. The signature of entanglement remains the same: the entropy reconstructed from the final-state hadrons should be equal to the entanglement entropy of the initial-state partons. 

\begin{figure}[thb]
\includegraphics[width=3.5in]{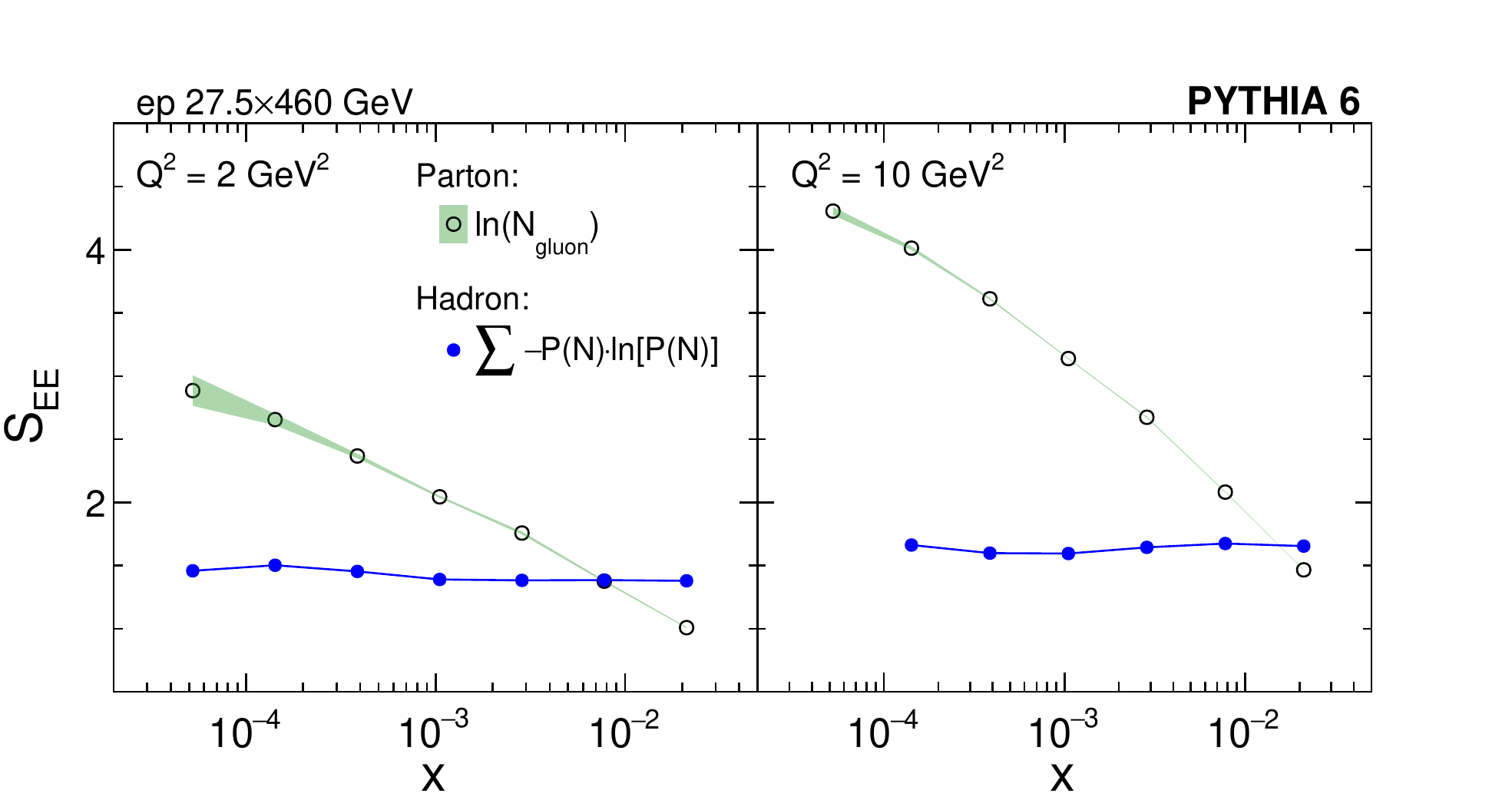}
  \caption{ \label{fig:figure_2} Entanglement entropy in Monte Carlo generator PYTHIA 6. The entanglement entropy predicted from the gluon distribution (MSTW) as a function of $x$ at momentum transfer scale $Q^{2}$ = 2 and 10 GeV$^{2}$ (open circles). The green band represents the symmetric systematic uncertainty at 90\% C.L. The entropy obtained from the final-state hadron multiplicity distribution, $\rm P(N)$, from PYTHIA 6 simulations at the same values of $Q^{2}$ is shown via blue filled circles. The statistical uncertainty is invisible on the scale of the plot.}
\end{figure}

\section{Results}

Let us begin by testing the proposed idea in the electron-proton DIS using Monte Carlo simulations. Since the probabilistic Monte Carlo event generators do not incorporate entanglement, we do not expect relationship (\ref{test}) to hold. Note that the available experimental data on hadron distributions in DIS, e.g., from Hadron-Electron Ring Accelerator (HERA) experiments, do not cover the kinematic regime of interest ($x<10^{-3}$) where relation (\ref{test}) applies.

First we obtain the the number of gluons, $\rm N_{gluon}$, by integrating the gluon distribution $xG(x)$ over a given $x$ range at a chosen scale $Q^{2}$. We use the leading order Parton Distribution Function (PDF) set MSTW at the 90\% C.L~\cite{Martin:2009iq}, shown in Fig.~\ref{fig:figure_6}. The entanglement entropy $\rm \ln{(N_{gluon})}$ predicted from the gluon distribution is shown in open black circles with systematic uncertainties depicted as green band in Fig.~\ref{fig:figure_2}. The entropy of the final-state hadrons is shown as blue filled circles. It is calculated from the multiplicity distribution, $\rm P(N)$, in a rapidity range determined by the $x$ range used to derive $\rm N_{gluon}$. For details see Appendix.~\ref{sec:appendix-1}. $\rm P(N)$ is taken from $ep$ DIS events created with the PYTHIA 6 event generator~\cite{Sjostrand:2006za}. 
We have tested several Monte Carlo event generators, such as PYTHIA 6, PYTHIA 8~\cite{Sjostrand:2014zea}, and DJANGO~\cite{Schuler:1991yg}, and have found that they give similar results. 
An example final-state hadron multiplicity distribution is shown in Fig.~\ref{fig:figure_7}. 

It becomes clear from Fig.~\ref{fig:figure_2}, that the two entropies, the von Neumann entanglement entropy associated with the gluon distribution, $\rm \ln{(N_{gluon})}$, and the entropy reconstructed from the final state hadrons, $S_\mathrm{hadron}$, are uncorrelated, as expected for Monte Carlo models that do not possess quantum entanglement. This correlation is absent for all MC generators that we have studied.

\begin{figure*}[th]
\centering
\includegraphics[width=6.7in]{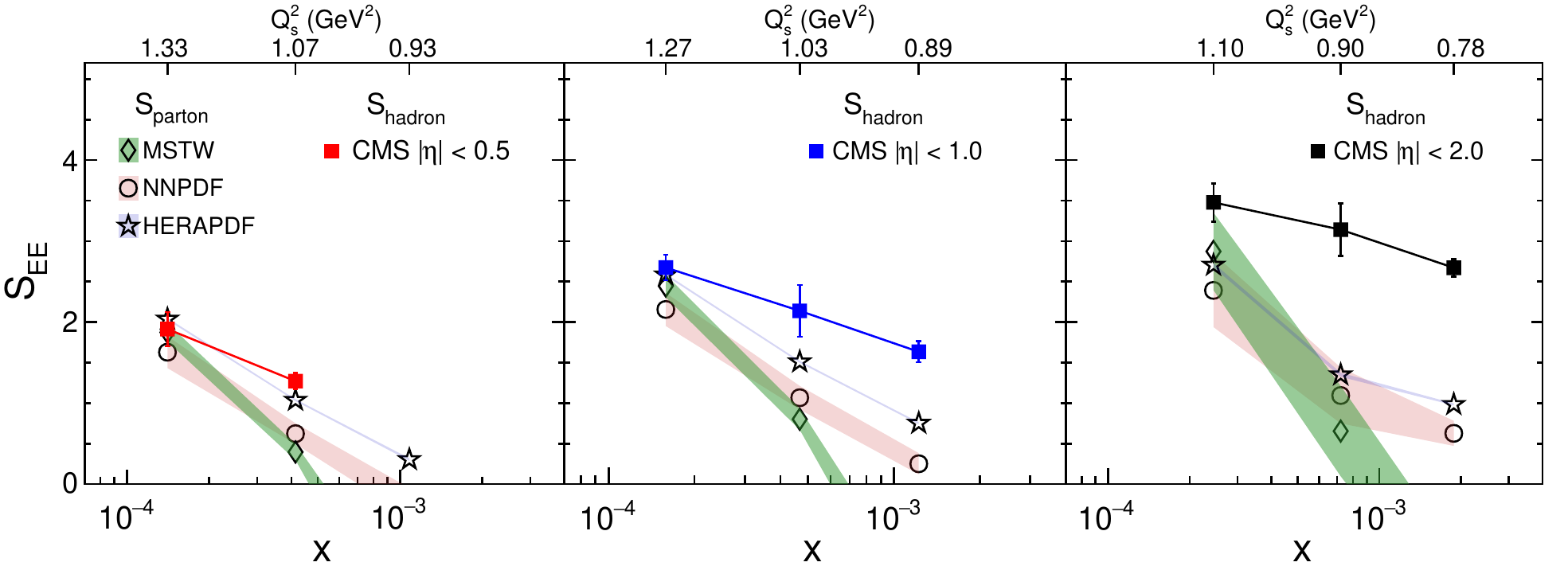}
   \caption{ \label{fig:figure_3} Entanglement entropy in proton-proton collisions at the LHC. The Boltzmann entropy calculated based on the final-state multiplicity $\rm P(N)$ distribution in $pp$ collisions~\cite{Khachatryan:2010nk} at the LHC in different pseudorapidity ranges is shown as a function of $x$ indicated by the filled squares, where the total uncertainty is denoted by the error bars. The expected entanglement entropies from the gluon distribution are presented with open markers using different PDF sets~\cite{Martin:2009iq,Ball:2012cx,Abramowicz:2015mha}. The bands denote the systematic uncertainty associated with the PDF. The saturation scale $Q^{2}_{s}$~\cite{Ducloue:2019ezk} for gluons at each $x$ value is indicated on the top axis. }
\end{figure*}

With a clearly drawn baseline from the Monte Carlo models, we can now look for entanglement in available experimental data. Since suitable data in $ep$ collisions do not exist, we have to turn for our study to $pp$ collisions using data from the CMS experiment~\cite{Khachatryan:2010nk} at the LHC. 
As outlined earlier the signature of entanglement stays the same (see also Fig.~\ref{fig:figure_1}).

 By performing an analysis similar to the one presented in Fig.~\ref{fig:figure_2}, we arrive at the results depicted in Fig.~\ref{fig:figure_3}. Here we show the comparison of the entanglement entropy predicted from the gluon distribution (three different leading order PDF sets are indicated by open symbols), and the Boltzmann entropy based on the final-state hadron multiplicity $\rm P(N)$ distribution (in filled symbols) as a function of $x$. Since $x$ and momentum transfer scale $Q^{2}$ are not directly available in $pp$ collisions (unlike in $ep$ experiments), an alternative way of comparing the entropy at similar $x$ and scales are used as detailed in Sec.~\ref{sec:appendix}. The experimental data from CMS are shown in three different pseudorapidity\footnote{pseudorapidity $\eta\equiv-\ln{(\tan{\frac{\theta}{2}})}$ and the $\theta$ is the angle with respect to the beam axis.} $\eta$ ranges within $\pm0.5$, $\pm1.0$, and $\pm2.0$ units. Data from pseudorapidity range within $|\eta|<1.5$ and 2.4 are depicted in Fig.~\ref{fig:figure_4} in Appendix.~\ref{sec:appendix-3}. The measurements of multiplicity distribution in $pp$ collisions performed by other LHC experiments, \textit{e.g.}, ATLAS~\cite{Aad:2016xww} and ALICE~\cite{ALICE:2017pcy}, are consistent with CMS for the same pseudorapidity ranges where available. In this analysis we use only CMS data since they provide 
 a wide set of central pseudorapidity ranges not available elsewhere.

In contrast with what we have observed in $ep$ Monte Carlo simulations, we find that 
the experimental data from $pp$ collisions at the LHC presented in Fig.~\ref{fig:figure_3} shows a striking agreement between the entanglement entropy predicted from the gluon distributions and the Boltzmann entropy from the final-state hadron multiplicity distributions in all $|\eta|$ ranges. The relation (\ref{test}) is expected~\cite{Kharzeev:2017qzs} to hold for $x<10^{-3}$, and the data is in good agreement with this prediction. This observation provides a strong direct indication of quantum entanglement at sub-nucleonic scales. The discrepancy observed towards higher $x$ might be due to non-negligible contributions from sea quarks at low $Q^{2}$ scales. A theoretical computation of the entanglement entropy including sea quarks is not yet available and will be an important study in the future.

\section{Summary}
The results reported in this letter provide a new perspective in understanding the proton structure and shed light on the nature of colour confinement. Our analysis of the experimental data from the LHC supports the resolution of an apparent paradox between the parton model and quantum mechanics based on quantum entanglement. In the future, it will be imperative to verify this result in electron-proton and electron-ion collisions at small $x$ that will require a future facility such as the Electron-Ion Collider~\cite{Accardi:2012qut}. It will be also important to study the real-time evolution of quantum entanglement in high energy processes - such studies have already begun 
\cite{Baker:2017wtt,Berges:2017hne,Kovner:2018rbf,Armesto:2019mna}.

\section{Acknowledgment}
We thank Stefan Schmitt for fruitful discussions on HERA experiments, and we thank Bertrand Ducloué, and Adrian Dumitru for providing their calculations of saturation scales. The work of Z.T is supported by LDRD-039 and the Goldhaber Distinguished Fellowship at Brookhaven National Laboratory. The Work of Z.T. and T.U. is supported by the U.S. Department of Energy under Award DE-SC0012704.
 The work of D.K. was supported in part by the U.S. Department of Energy under Awards DE-FG-88ER40388 and DE-AC02-98CH10886.

\appendix
\section{\label{sec:appendix} Appendix}

\subsection{\label{sec:appendix-1} Electron-proton collisions} 
The DIS process, $e + p \rightarrow e^\prime + X$, proceeds through the exchange of a virtual photon between the electron and the proton. The scattering process can be characterized in terms of $x$ and $Q^2$ that can be derived by measuring the energy and angle of the scattered electron $e^\prime$. In order to calculate $\rm N_{gluon}$ we integrate the known gluon distribution 
$xG(x)$ over a given $x$ range for a fixed $Q^2$. $\rm N_{gluon}$ then defines the entanglement entropy $\rm \ln{(N_{gluon})}$ which is then compared to the Boltzmann entropy, \textbf{$\rm -\sum{P(N)\cdot\ln{P(N)}}$}, derived from
the final-state hadrons. The phase space in which the final-state hadrons are measured is related to the $x$ range over which we integrate $xG(x)$ by $\ln{(1/x)}\approx\rm y_{_\mathrm{proton}}-y_{_\mathrm{hadron}}$~\cite{lecture_McLerran}, where $y_{_\mathrm{proton}}$ is the proton beam rapidity\footnote{rapidity $y\equiv\frac{1}{2}\ln{\frac{E+p_{z}c}{E-p_{z}c}}$. $E$ and $p_{z}$ are the energy and the longitudinal momentum of the particle.} and $y_{_\mathrm{hadron}}$ is the final-state hadron rapidity. For example, events with 27.5 GeV electrons scattering off 460 GeV protons with $x$ between $3\times10^{-5}$ and $8\times10^{-5}$ correspond to a rapidity range of $-3.5<y<-2.5$. It is in this rapidity range where the final-state hadron multiplicity $\rm P(N)$ distribution is measured and it is the respective $x$ range that is used to integrate over $xG(x)$.

\subsection{\label{sec:appendix-2} Proton-proton collisions} 

For $pp$ collision, $p + p \rightarrow X$, the final-state hadron multiplicity $\rm P(N)$ distributions are taken from the published result of the CMS experiment~\cite{Khachatryan:2010nk} at the LHC using $pp$ minimum-bias inelastic collisions at center-of-mass energies of 7, 2.36, and 0.9 TeV. Here we adopt the same relation as in $ep$, namely $\ln{(1/x)}\approx y_{_\mathrm {proton}}-y_{_\mathrm {hadron}}$ to calculate $x$ that cannot be directly measured in $pp$. At the same rapidity, a collision with higher center-of-mass energy corresponds to a lower $x$. At each energy, the data in different pseudorapidity ranges are depicted in Fig.~\ref{fig:figure_3}. Each range corresponds to different range in $x$ that is used to integrate over the gluon distributions $xG(x)$ to derive $\rm N_{gluon}$. CMS presents their data in intervals of pseudorapidity not rapidity. We included the uncertainties caused by the differences in the error bars of the CMS data in Fig.~\ref{fig:figure_3}, which are around $\sim4$--5\% estimated from PYTHIA simulations. Currently the CMS data spans a $x$ range from $\sim10^{-4}$ to $\sim10^{-2}$. 

A conceptual problem in $pp$ collisions is that two gluon distributions are involved, one from each proton, while we calculate the
entanglement entropy from one distribution. Instead of altering the definition of the entanglement entropy, which would introduce strong model dependencies, we modify the $\rm P(N)$ distributions by extrapolating the $\rm P(N)$ distribution to reflect a single proton similar to that in DIS experiment. We do so by fitting a generalized Negative Binomial Distribution (NBD)~\cite{Khachatryan:2010nk} to the $\rm P(N)$ distributions. The final $\rm P(N)$ is then taken as the same NBD function
but with only half of the average multiplicity. 
This approach relies on the assumption that the final-state hadrons are produced coherently by the two colliding protons instead by incoherent and independent fragmentation. 
Different NBD fit ranges and different fit functions, such as double NBDs, were also conducted. The difference with respect to the default result is folded into the error bars presented in Fig.~\ref{fig:figure_3}.

\begin{figure}[thb]
\includegraphics[width=2.8in]{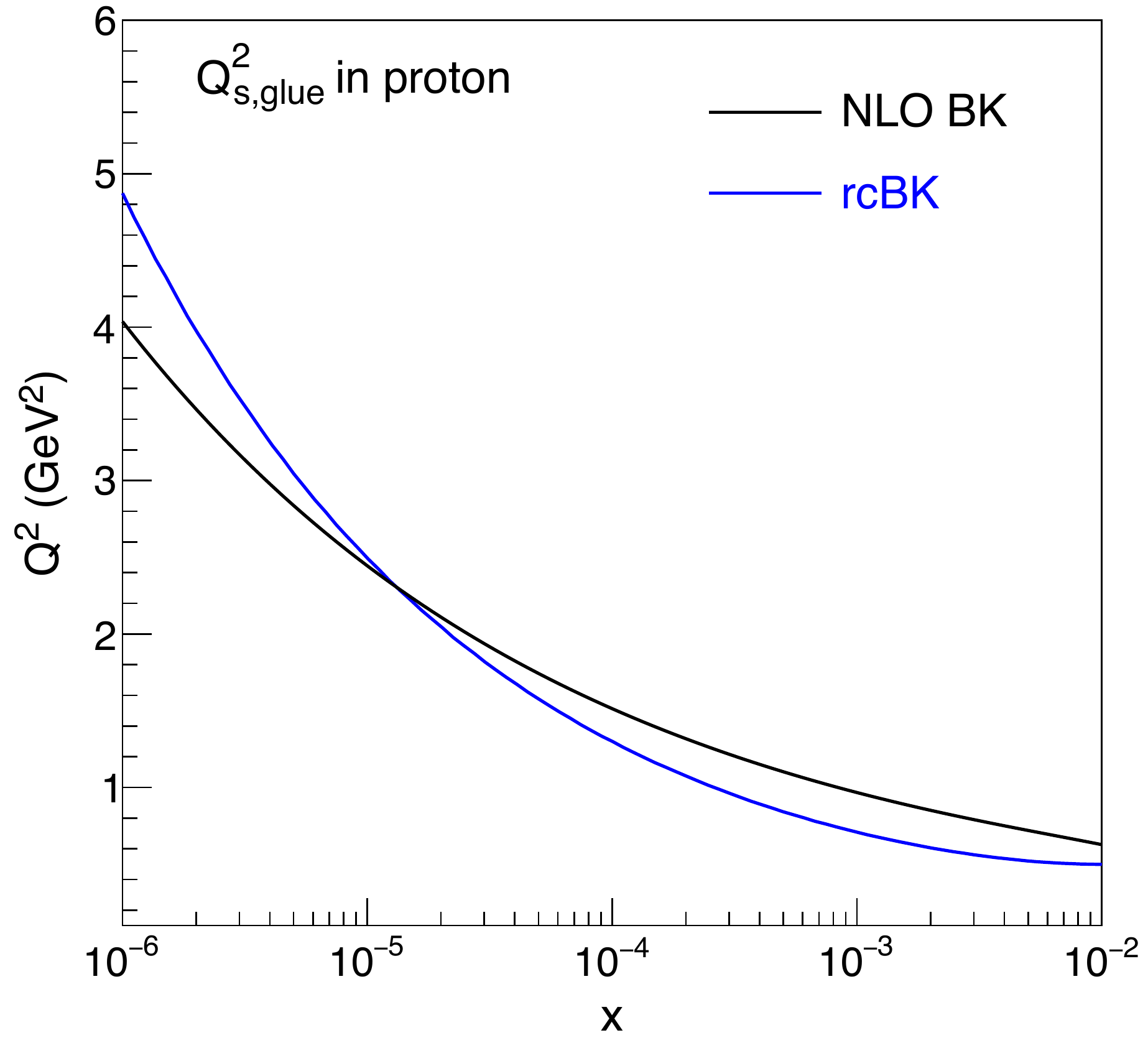}
  \caption{ \label{fig:figure_5} Saturation scale $Q^{2}_{s}$ as a function of $x$. The saturation scale $Q^{2}_{s}$ as predicted by rcBK~\cite{Albacete:2012xq} (blue curve)  and NLO BK~\cite{PhysRevC.85.034907} (black curve) calculations are shown as a function of $x$. For our main result we used NLO BK. }
\end{figure}

In hard $pp$ collisions accompanied by the production of jets, the appropriate scale $Q^2$ is set by the transverse momentum of the jet. In an average $pp$ collision, the scale is set instead by a characteristic transverse momentum of partons in the proton's wave function. This momentum is determined by the density of partons in the transverse plane; since the density of partons saturates at small $x$, the corresponding scale is referred to as the``saturation momentum" $Q_s$. 
For determining the entanglement entropy from $\ln\left[xG(x)\right]$ we use the saturation scale $Q^{2}_{s}(x)$ derived in NLO BK calculations~\cite{PhysRevC.85.034907} (indicated by the black curve in Fig.~\ref{fig:figure_5}), which reasonably reproduces particle production at the LHC. For each $x$, the corresponding $Q^{2}_{s}(x)$ value is indicated on the top axes of Fig.~\ref{fig:figure_3}. An alternative rcBK~\cite{Albacete:2012xq} calculation of $Q^{2}_{s}(x)$ is depicted together with the NLO BK model in Fig.~\ref{fig:figure_5}. Both represent the most recent calculation of $Q^{2}_{s}(x)$ and give very similar results in the $x$ range studied here. A complete summary of $x$ and $Q^{2}_{s}$ values is given for all pseudorapidity ranges in Tables~\ref{tab:rapidity_0p5}-\ref{tab:rapidity_2p4}.

\clearpage
\subsection{\label{sec:appendix-3} Supplemental figures and tables}

\begin{figure}[thb]
\includegraphics[width=3.5in]{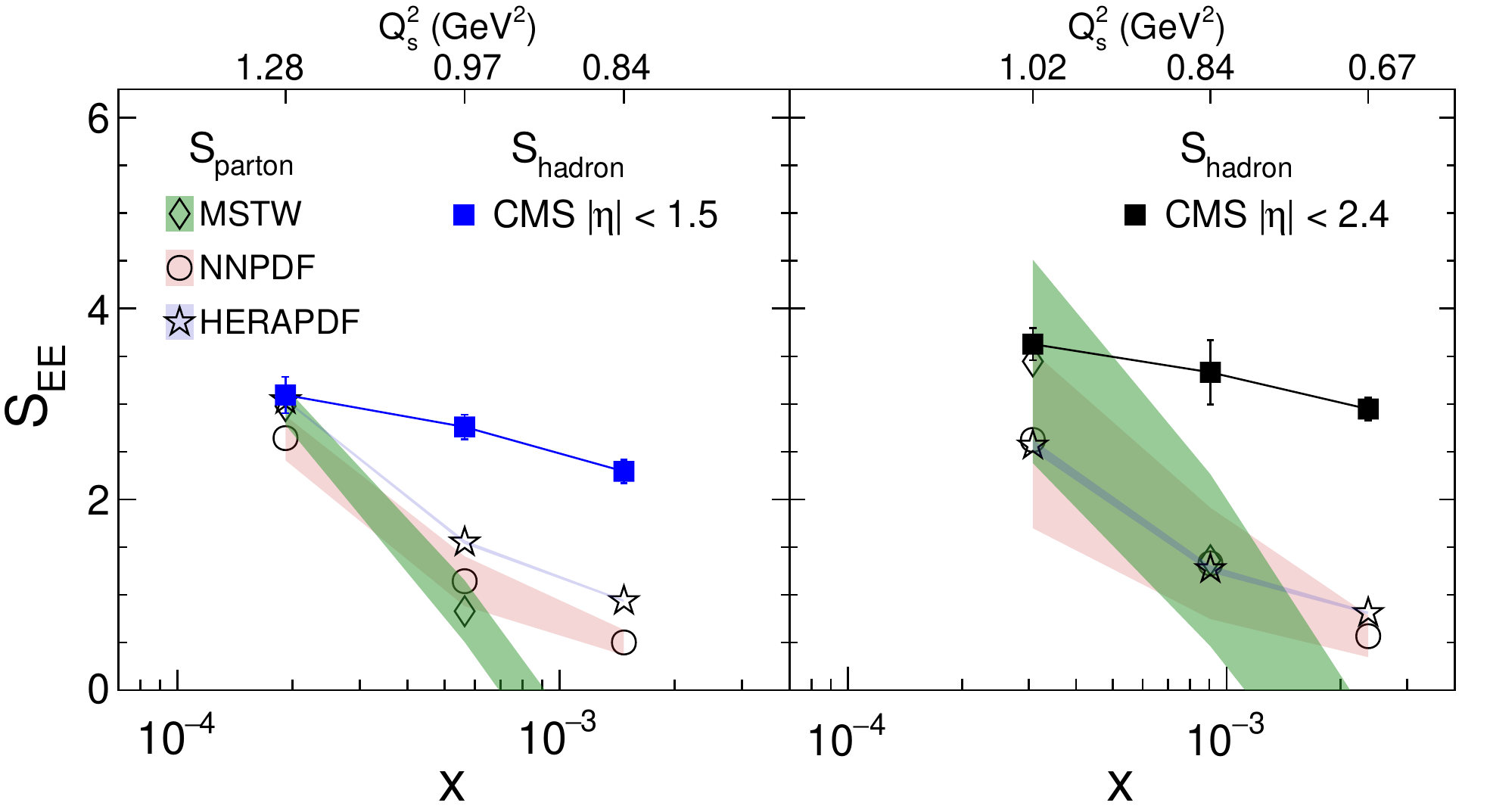}
  \caption{ \label{fig:figure_4} Entanglement entropy in proton-proton collisions at the LHC. The Boltzmann entropy calculated based on the final-state multiplicity $\rm P(N)$ distribution in $pp$ collisions at the LHC are shown as a function of $x$ indicated by the filled squares, where the total uncertainty is denoted by the error bars. The selected pseudorapidity ranges are within $|\eta|<1.5$ and 2.4. The expected entanglement entropies from the gluon distribution are presented with open markers using different PDF sets for comparison. The bands denote the systematic uncertainty from the PDF sets. The saturation scale $Q^{2}_{s}$ for gluon at each x value is indicated on the top axis. We used $Q^{2}_{s}$ derived in a recent NLO BK calculation~\cite{PhysRevC.85.034907}. }
\end{figure}

\begin{figure}[thb]
\includegraphics[width=3in]{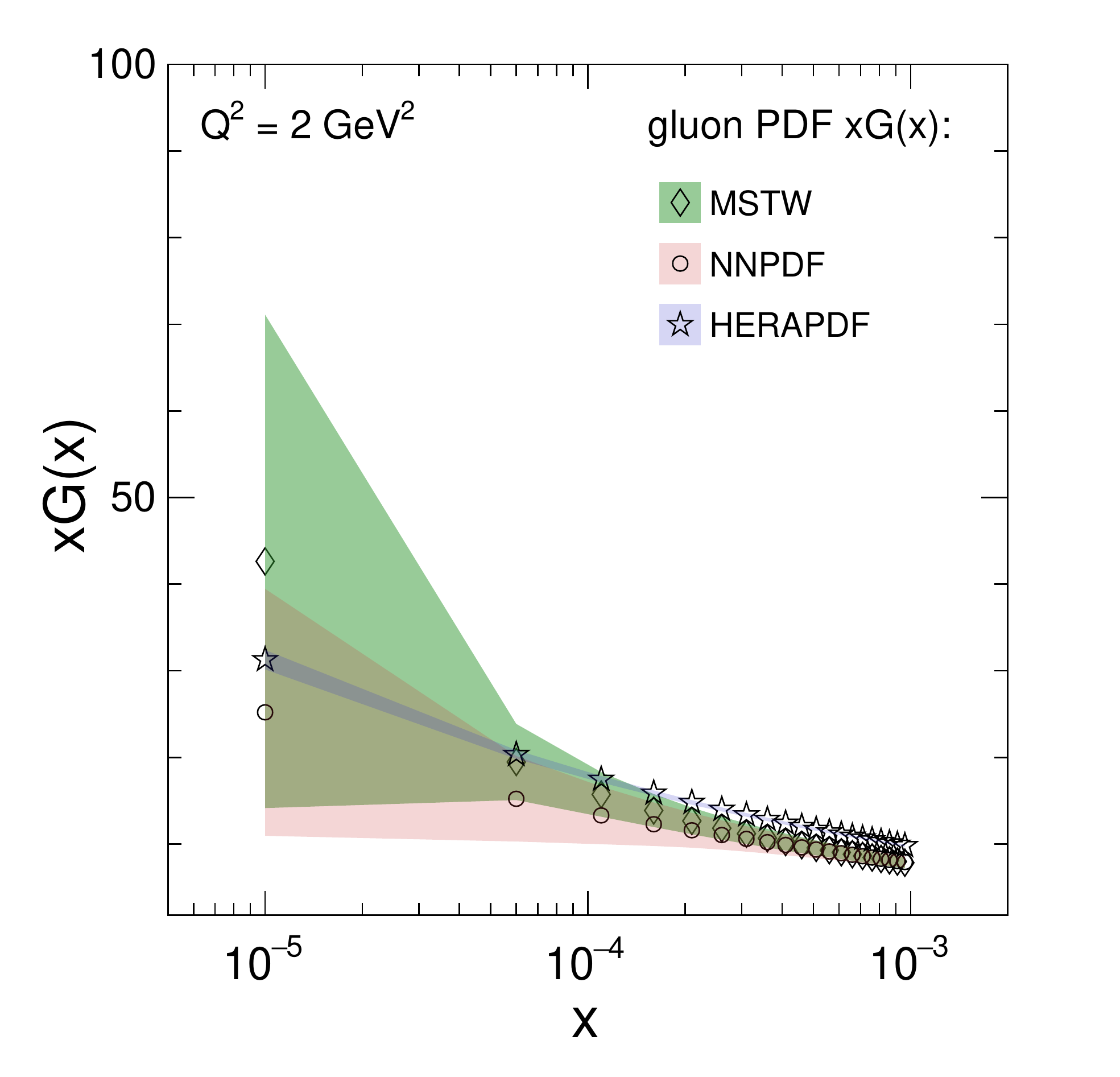}
  \caption{ \label{fig:figure_6} Gluon distributions. The Parton Distribution Functions (PDF) for gluons are shown as a function of $x$ at the momentum transfer scale $Q^{2}=2\ \mathrm{GeV}^{2}$. Three different leading order PDF sets are presented, {MSTW}~\cite{Martin:2009iq}, {NNPDF}~\cite{Ball:2012cx}, and {HERAPDF}~\cite{Abramowicz:2015mha}. The systematic uncertainty are shown as the coloured bands, suggested by the respective PDF authors. }
\end{figure}

\begin{figure}[thb]
\includegraphics[width=3in]{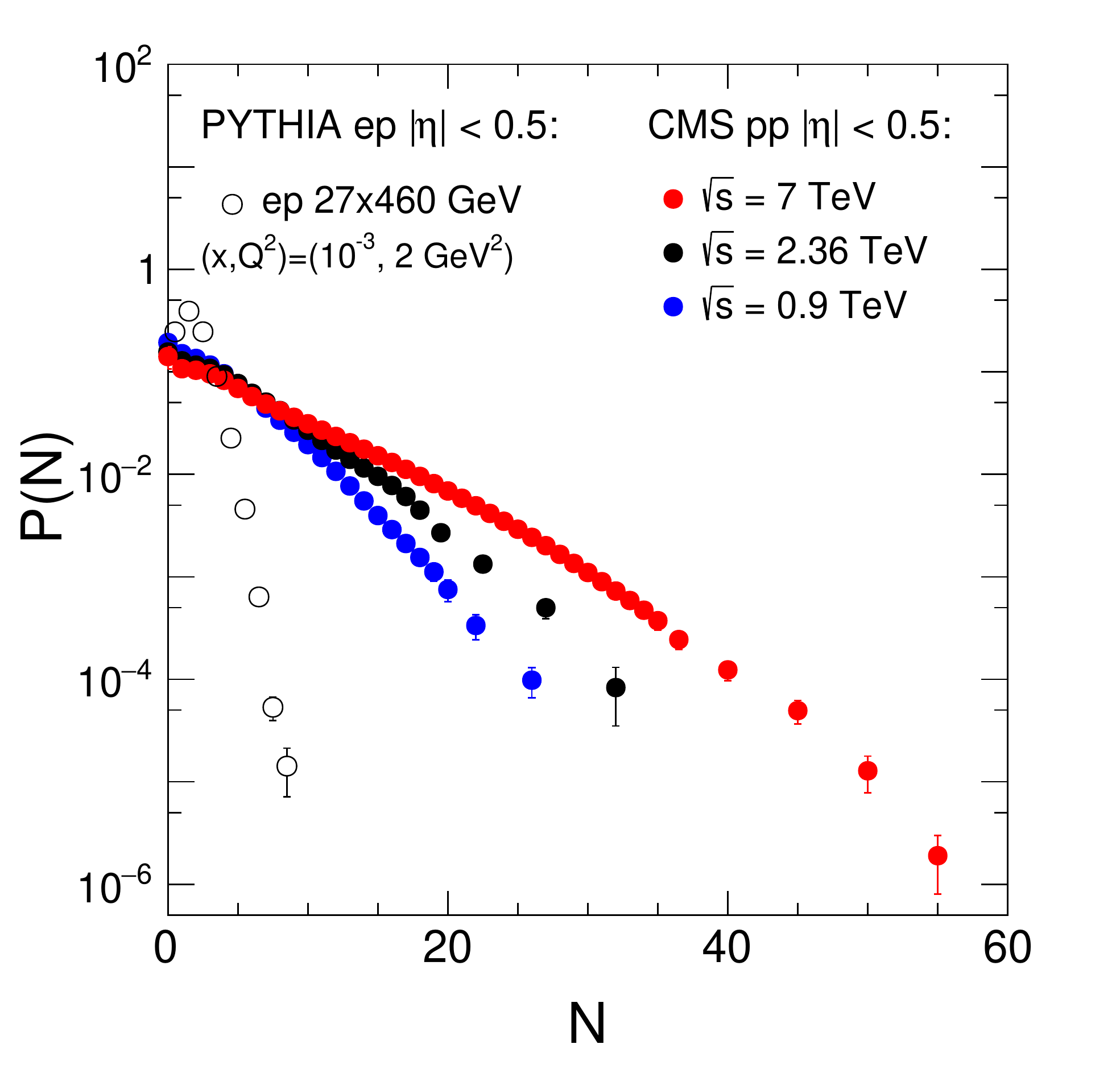}
  \caption{ \label{fig:figure_7} Final-state hadron multiplicity distributions. The final-state hadron multiplicity distributions, $\rm P(N)$, in electron-proton collisions from PYTHIA 6 simulation and $pp$ collisions from the CMS experiment~\cite{Khachatryan:2010nk} at various of center-of-mass energies (7, 2.37, and 0.9 TeV), are shown as a function of N particles per event. Here, final-state hadron were selected within the pseudorapidity range $|\eta|<0.5$ and transverse momentum greater than zero.}
\end{figure}

\begin{table*}
\caption{\label{tab:rapidity_0p5}} Summary of kinematic variables in $pp$ collisions. 
This table summarizes values of $x$ and the saturation scale $Q^{2}_{s}$ based on the final-state hadron and proton beam rapidity. The hadron rapidity is selected within $|y|<0.5$. The relation between rapidity $y$ and momentum fraction $x$ is given by $\ln{\left(1/x_{i}\right)}\simeq y_\mathrm{beam} - y_\mathrm{hadron,i}$. The subscript 1 and 2 denote the boundary value according to their range for the given rapidity interval. 
\begin{tabular}{cccccccc}
\hline
\hline
 CM energy $\sqrt{s}$ & $y_\mathrm{beam}$ & $x_{1}$ & $x_{2}$ & $\left\langle x \right\rangle $ & $Q^{2}_{s,1}$ & $Q^{2}_{s,2}$ & $\left\langle Q^{2}_{s} \right\rangle$ \\
 \hline
 7 TeV & 8.91 &  8.13E-05 & 2.21E-04 & 1.41E-04 & 1.57E+00 & 1.31E+00 & 1.33E+00 \\
 2.36 TeV & 7.83 &  2.41E-04 & 6.55E-04 & 4.14E-04 & 1.26E+00 & 1.06E+00 & 1.07E+00 \\
 0.9 TeV & 6.87 &  6.32E-04 & 1.72E-03 & 1.08E-03 & 1.04E+00 & 8.80E-01 & 9.30E-01 \\

 \hline
 \end{tabular}
 \end{table*}

 \begin{table*}
\caption{\label{tab:rapidity_1p0}} Summary of kinematic variables in $pp$ collisions. This table summarizes values of $x$ and saturation scale $Q^{2}_{s}$ based on the final-state hadron rapidity and the proton beam rapidity. The hadron rapidity is selected within $|y|<1.0$. The relation between rapidity $y$ and momentum fraction $x$ is given by $ \ln{\left(1/x_{i}\right)}\simeq y_\mathrm{beam} - y_\mathrm{hadron,i}$. The subscript 1 and 2 denote the boundary value according to their range for the given rapidity interval. 

\begin{center}
\begin{tabular}{cccccccc}
\hline
\hline
 CM energy $\sqrt{s}$ & $\rm y_\mathrm{beam}$ & $x_{1}$ & $x_{2}$ & $\left\langle x \right\rangle $ & $Q^{2}_{s,1}$ & $ Q^{2}_{s,2}$ & $\left\langle Q^{2}_{s} \right\rangle$ \\
 \hline
 7 TeV & 8.91 &  4.93E-05 & 3.64E-04 & 1.59E-04 & 1.74E+00 & 1.19E+00 & 1.27E+00 \\
 2.36 TeV & 7.83 &  1.46E-04 & 1.08E-03 & 4.67E-04 & 1.39E+00 & 9.60E-01 & 1.03E+00 \\
 0.9 TeV & 6.87 &  3.83E-04 & 2.83E-03 & 1.22E-03 & 1.14E+00 & 8.00E-01 & 8.90E-01 \\

 \hline
 \end{tabular}
 \end{center}
 \end{table*}

\begin{table*}
\caption{\label{tab:rapidity_1p5}} Summary of kinematic variables in $pp$ collisions. This table summarizes values of $x$ and the saturation scale $Q^{2}_{s}$ based on the final-state hadron rapidity and the proton beam rapidity. The hadron rapidity is selected within $|y|<1.5$. The relation between rapidity $y$ and momentum fraction $x$ is given by 
$ \ln{\left(1/x_{i}\right)}\simeq y_\mathrm{beam} - y_\mathrm{hadron,i}$. The subscript 1 and 2 denote the boundary value according to their range for the given rapidity interval. 

\begin{center}
\begin{tabular}{cccccccc}
\hline
\hline
 CM energy $\sqrt{s}$ & $ y_\mathrm{beam}$ & $x_{1}$ & $x_{2}$ & $\left\langle x \right\rangle $ & $Q^{2}_{s,1}$ & $ Q^{2}_{s,2}$ & $\left\langle Q^{2}_{s} \right\rangle$ \\
 \hline
 7 TeV & 8.91 &   2.99E-05 & 6.01E-04 & 1.92E-04 & 1.93E+00 & 1.06E+00 & 1.28E+00\\
 2.36 TeV & 7.83 &   8.87E-05 & 1.78E-03 & 5.64E-04 & 1.54E+00 & 8.80E-01 & 9.70E-01\\
 0.9 TeV & 6.87 &   2.33E-04 & 4.67E-03 & 1.47E-03 & 1.26E+00 & 7.30E-01 & 8.40E-01\\

 \hline
 \end{tabular}
 \end{center}
 \end{table*}

\begin{table*}
\caption{\label{tab:rapidity_2p0}} Summary of kinematic variables in $pp$ collisions. This table summarizes values of $x$ and the saturation scale $Q^{2}_{s}$ based on the final-state hadron rapidity and the proton beam rapidity. The hadron rapidity is selected within $|y|<2.0$. The relation between rapidity $y$ and momentum fraction $x$ is given by 
$\ln{\left(1/x_{i}\right)}\simeq y_\mathrm{beam} - y_\mathrm{hadron,i}$. The subscript 1 and 2 denote the boundary value according to their range for the given rapidity interval.

\begin{center}
\begin{tabular}{cccccccc}
\hline
\hline
 CM energy $\sqrt{s}$ & $y_\mathrm{beam}$ & $x_{1}$ & $x_{2}$ & $\left\langle x \right\rangle $ & $Q^{2}_{s,1}$ & $ Q^{2}_{s,2}$ & $\left\langle Q^{2}_{s} \right\rangle$ \\
 \hline
 7 TeV & 8.91 &   1.81E-05 & 9.90E-04 & 2.45E-04 & 2.14E+00 & 9.80E-01 & 1.10E+00\\
 2.36 TeV & 7.83 & 5.38E-05 & 2.94E-03 & 7.21E-04 & 1.70E+00 & 8.00E-01 & 9.00E-01  \\
 0.9 TeV & 6.87 &  1.41E-04 & 7.70E-03 & 1.88E-03 & 1.39E+00 & 6.60E-01 & 7.80E-01 \\

 \hline
 \end{tabular}
 \end{center}
 \end{table*}

\begin{table*}
\caption{\label{tab:rapidity_2p4}} Summary of kinematic variables in $pp$ collisions. This table summarizes values of $x$ and the saturation scale $Q^{2}_{s}$ based on the final-state hadron rapidity and the proton beam rapidity. The hadron rapidity is selected within $|y|<2.4$. The relation between rapidity $y$ and momentum fraction $x$ is given by 
$\ln{\left(1/x_{i}\right)}\simeq y_\mathrm{beam} - y_\mathrm{hadron,i}$. The subscript 1 and 2 denote the boundary value according to their range for the given rapidity interval. 

\begin{center}
\begin{tabular}{cccccccc}
\hline
\hline
 CM energy $\sqrt{s}$ & $y_\mathrm{beam}$ & $x_{1}$ & $x_{2}$ & $\left\langle x \right\rangle $ & $Q^{2}_{s,1}$ & $ Q^{2}_{s,2}$ & $\left\langle Q^{2}_{s} \right\rangle$ \\
 \hline
 7 TeV & 8.91 &   1.22E-05 & 1.48E-03 & 3.08E-04 & 2.33E+00 & 9.10E-01 & 1.02E+00 \\
 2.36 TeV & 7.83 &   3.61E-05 & 4.38E-03 & 9.06E-04 & 1.85E+00 & 7.40E-01 & 8.40E-01 \\
 0.9 TeV & 6.87 &   9.45E-05 & 1.15E-02 & 2.37E-03 & 1.51E+00 & 6.20E-01 & 6.70E-01 \\

 \hline
 \end{tabular}
 \end{center}
 \end{table*}
 
 \clearpage

\bibliography{manuscript_ee}% Produces the bibliography via BibTeX.

\end{document}